\documentclass[a4paper,12pt]{article}
\usepackage{amssymb,amsthm,latexsym}

\newcommand{\med}[1]{\langle #1 \rangle}
\newtheorem{theorem}{Theorem}
\newtheorem{lemma}{Lemma}
\newtheorem{definition}{Definition}

\title{Broken Replica Symmetry Bounds in the Mean Field Spin Glass Model}

\author{
Francesco Guerra\footnote{\
e-mail: {\tt francesco.guerra@roma1.infn.it}} \\
{\small {\itshape Dipartimento di Fisica, Universit\`a di Roma ``La Sapienza''}}
\\
{\small {\itshape and INFN, Sezione di Roma1, Piazzale A. Moro 2, 00185 Roma, 
Italy}}
} 

\date{\today}
\begin{document}
\maketitle
\begin{abstract}
By using a simple interpolation argument, in previous work we have proven 
the existence of the thermodynamic limit, for mean field disordered models, 
including the Sherrington-Kirkpatrick model, and the Derrida p-spin model. 
Here we extend this argument in order to compare the limiting free energy 
with the expression given by the Parisi {\it Ansatz}, and including full 
spontaneous replica symmetry breaking. Our main result is that the quenched 
average of the free energy is bounded from below by the value given in the Parisi 
{\it Ansatz}, uniformly in the size of the system. Moreover, the difference
between the two expressions is given in the form of a sum rule, extending 
our previous work on the comparison between the true free energy and its 
replica symmetric Sherrington-Kirkpatrick approximation. We give also a 
variational bound for the infinite volume limit of the ground state energy 
per site. 
\end{abstract}
\newpage
\section{Introduction}
\label{intro}

The main objective of this paper is to compare the free energy of the mean 
field spin glass model, introduced by Sherrington and Kirkpatrick in \cite{SK}, 
with the expression given in 
the frame of the Parisi {\it Ansatz} \cite{P}, \cite{MPV}, including 
the complete phenomenon of spontaneous replica symmetry breaking. In 
previous work \cite{Gsum}, we have limited our comparison to the replica 
symmetric case, by producing sum rules, where the difference between the 
true free energy, and its replica symmetric approximation, is expressed in 
terms of overlap fluctuations, with a well definite sign. As a result, the 
replica symmetric approximation turns out to be a rigorous lower bound for 
the quenched average of the free energy per site, uniformly in the size of 
the system. 

In the meantime, the old problem of proving the existence of 
the infinite volume limit for the thermodynamic quantities has been solved 
\cite{GTthermo}, by using a simple comparison argument.

Here, we extend this comparison argument, by introducing an appropriate 
generalized partition function, as a function of a parameter $t$, 
with $0 \le t \le 1$, able to interpolate between the true theory, at 
$t=1$, and the broken replica  {\it Ansatz}, at $t=0$. Consequently, through 
a simple direct calculation, we can evaluate the difference between the 
true free energy, and its broken replica  expression, still in the form of a 
sum rule, with the corrections, of a definite sign, expressed through overlap 
fluctuations, in properly chosen auxiliary states. As a result, the broken replica 
 {\it Ansatz} turns out to be a rigorous lower bound for the 
quenched average of the free energy per site, uniformly in the size of the 
system.

Moreover, the corrections, given in terms of overlap fluctuations, are in 
a form suitable for the exploration of the expected result of their vanishing, 
when the size of the system goes to infinite, along the program developed 
in \cite{GTquadr}.

Of course, our method does not use the replica trick in the zero replica 
limit, as explained for example in \cite{MPV}, nor the cavity method, as 
exploited for example in \cite{MPVcavity}, \cite{PS}, \cite{Glocarno}, 
\cite{Tbook}.

We give only a brief sketch of the extension of our method to the Derrida 
p-spin model \cite{D}, \cite{GM}, \cite{Ga}, \cite{Tbook}. A more 
detailed treatment will be presented elsewhere \cite{GTpspin}.

The organization of the paper is as follows. In Section 2, we will briefly 
recall the main features, and definitions, of the mean field spin glass 
model. In Section 3, the general structure of the Parisi spontaneously 
broken replica symmetry {\it Ansatz} will be described, in a form suitable 
for the developments of  next Section. Section 4 contains the main results 
of the paper. Firstly, we introduce the interpolating generalized 
partition function. Then, we evaluate its derivative, with respect to the 
interpolating parameter, arriving to the sum rule. The general broken replica 
 bound follows easily. In Section 5, we give a variational estimate 
for the infinite volume limit of the ground state enrgy per site. 
Next Section 6 gives the essential 
ingredients of the extension of this method to the p-spin model. Finally, 
Section 7 is devoted to conclusions and outlook for further developments.

\section{The basic definitions for the mean field spin glass model}

The generic configuration of the mean field spin glass model is defined 
through Ising spin variables
$\sigma_{i}=\pm 1$,  attached to each site $i=1,2,\dots,N$. 

The external quenched disorder is
given by the $N(N-1)/2$ independent and identical distributed random
variables $J_{ij}$, defined for each couple of sites. For the sake of simplicity,
we assume each $J_{ij}$ to be a centered
unit Gaussian with averages $E(J_{ij})=0$, $ E(J_{ij}^2)=1$.

The Hamiltonian of the model, in some external field of strength $h$,  
is given by the mean field expression
\begin{equation}\label{H}
H_N(\sigma,h,J)=-{1\over\sqrt{N}}\sum_{(i,j)}J_{ij}\sigma_i\sigma_j
-h\sum_{i}\sigma_i.
\end{equation}
Here, the first sum extends to all site couples, an the second to all sites.

For a given inverse temperature $\beta$, let us now introduce the 
disorder dependent
partition function $Z_{N}(\beta,h,J)$, 
the quenched average of the free energy per site
$f_{N}(\beta,h)$, the Boltzmann state 
$\omega_J$, and the auxiliary function $\alpha_N(\beta,h)$,  
according to the well known definitions
\begin{eqnarray}\label{Z}
&&Z_N(\beta,h,J)=\sum_{\sigma_1\dots\sigma_N}\exp(-\beta H_N(\sigma,h,J)),\\
\label{f}
&&-\beta f_N(\beta,h)=N^{-1} E\log Z_N(\beta,h,J)=\alpha_N(\beta,h),\\
\label{state}
&&\omega_{J}(A)=Z_N(\beta,h,J)^{-1}\sum_{\sigma_1\dots\sigma_N}A\exp(-\beta
H_N(\sigma,h,J)), 
\end{eqnarray}
where $A$ is a generic function of the $\sigma$'s. 

Replicas are introduced by 
considering a generic number $s$ of independent copies
of the system, characterized by the Boltzmann
variables $\sigma^{(1)}_i$, $\sigma^{(2)}_i$, $\dots$,
distributed according to the product state 
$\Omega_J=\omega^{(1)}_J \omega^{(2)}_J \dots\omega^{(s)}_J$. Here,
all $\omega^{(\alpha)}_J$ act on each one
$\sigma^{(\alpha)}_i$'s, and are subject to the {\sl
same} sample $J$ of the external noise. 

The overlap between  two replicas $a,b$ is
defined according to
$q_{ab}=
N^{-1}\sum_{i}\sigma^{(a)}_i\sigma^{(b)}_i,$
with the obvious bounds
$-1\le q_{ab}\le 1$.

For a generic smooth function $F$ of the overlaps, we
define the $\langle\,\,\rangle$ averages
\begin{equation}
\label{medie}
\langle F(q_{12},q_{13},\dots)\rangle=
E\Omega_J\bigl(F(q_{12},q_{13},\dots)\bigr),
\end{equation}
where the Boltzmann averages $\Omega_J$ act on the replicated $\sigma$
variables, and $E$ is the average with respect to the external noise $J$.

\section{The broken replica symmetry  {\bf \it Ansatz}}

While we refer to the original paper \cite{P}, and to the extensive 
review given in \cite{MPV}, for the general motivations, and the 
derivation of the broken replica  {\it Ansatz}, in the frame of the 
ingenious replica trick, here we limit ourselves to a synthetic 
description of its general structure, in a form suitable for the treatment 
of next Section, see also \cite{Glocarno}, \cite{BR}.

First of all, let us introduce the convex space ${\cal X}$ of the functional 
order parameters $x$, as nondecreasing functions of the auxiliary variable 
$q$,
both $x$ and $q$ taking
values on the interval $[0,1]$, {\it i.e.}
\begin{equation}
\label{x}
{\cal X}\ni x : [0,1]\ni q \rightarrow x(q) \in [0,1].
\end{equation}
Notice that we call $x$ the nondecreasing function, and $x(q)$ its values.
We introduce a metric on ${\cal X}$ through the $L^{1}([0,1], dq)$ norm, where 
$dq$ is the Lebesgue measure.

Usually, we will consider the case of piecewise constant functional order 
parameters, characterized by an integer $K$, and two sequences $q_0, q_1, 
\dots, q_K$, $m_1, m_2, \dots, m_K$ of numbers satisfying
\begin{equation}
\label{qm}
0=q_0\le q_1 \le \dots \le q_{K-1} \le q_K=1,\,\,\, 0\le m_1 \le m_2 \le \dots 
\le m_K \le 1,
\end{equation}
such that
\begin{eqnarray}
\nonumber
x(q)=m_1 \,\,\mbox{for}\,\, 0=q_0\le q < q_1,\,\,\, 
x(q)=m_2 &\mbox{for}& q_1\le q < q_2,\\
\label{xpcws}
\ldots,
 x(q)=m_K &\mbox{for}& q_{K-1}\le q \le q_K.
\end{eqnarray}
In the following, we will find convenient to define also $m_0\equiv 0$, 
and $m_{K+1}\equiv 1$. The replica symmetric case corresponds to 
\begin{equation}
\label{replicas}
K=2,\,\, 
q_1=\bar q, \,\, m_1=0,\,\, m_2=1.
\end{equation}
The case $K=3$ is the first level of 
replica symmetry breaking, and so on.

Let us now introduce the function $f$, with values $f(q,y;x,\beta)$, of 
the variables $q\in[0,1]$, $y\in R$, depending also on the functional order 
parameter $x$, and on the inverse temperature $\beta$, defined as the 
solution of the nonlinear antiparabolic equation 
\begin{equation}
\label{antipara}
(\partial_q f)(q,y)+
{1\over2}\bigl(f^{\prime\prime}(q,y)+x(q){f^\prime}^2(q,y)\bigr)=0,
\end{equation}
with final condition
\begin{equation}
\label{final}
f(1,y)=\log\cosh(\beta y).
\end{equation}
Here, we have stressed only the dependence of $f$ on $q$ and $y$, and have 
put $f^\prime=\partial_y f$ and $f^{\prime\prime}=\partial_y^2 f$.

It is very simple to integrate Eq.~(\ref{antipara}) when $x$ is piecewise 
constant. In fact, consider $x(q)=m_a$, for $q_{a-1}\le q \le q_{a}$, 
firstly with $m_a>0$. Then, 
it is immediately seen that the correct solution of Eq.~(\ref{antipara}) in 
this interval, with the right final boundary condition at $q = q_{a}$, is 
given by
\begin{equation}
\label{solution}
f(q,y)=\frac{1}{m_a}\log \int\exp\bigl({m_{a} f(q_a,y+z\sqrt{q_a-q})}\bigr)\,d\mu(z),
\end{equation}
where $d\mu(z)$ is the centered unit Gaussian measure on the real line. On 
the other hand, if $m_a=0$, then (\ref{antipara}) loses the nonlinear part 
and the solution is given by
\begin{equation}
\label{solution0}
f(q,y)= \int f(q_a,y+z\sqrt{q_a-q})\,d\mu(z),
\end{equation}
which can be seen also as deriving from (\ref{solution}) in the limit $m_a \to 0$.
Starting from the last interval $K$, and using (\ref{solution}) iteratively on 
each interval, we easily get the solution of (\ref{antipara}), 
(\ref{final}), in the case of piecewise order parameter $x$, as in
(\ref{xpcws}).

We refer to \cite{Greview} for a detailed discussion about 
the properties of the solution
$f(q,y;x,\beta)$
of the antiparabolic equation (\ref{antipara}), with final condition 
(\ref{final}), as a
functional of a generic given $x$, as in (\ref{xpcws}). Here we only state 
the following
\begin{theorem}
The function $f$ is monotone in $x$, in the sense that 
$x(q)\le{\bar x}(q)$, for all $0\le q\le1$, implies 
$f(q,y;x,\beta)\le f(q,y;{\bar x},\beta)$, for any $0\le q\le1$, $y\in R$. 
Moreover $f$
is pointwise continuous in the $L^1([0,1],dq)$ norm. 
In fact, for generic $x$, $\bar x$, we
have    
$$|f(q,y;x,\beta)-f(q,y;{\bar x},\beta)|\le
{\beta^2\over2}\int^1_q|x(q')-{\bar x}(q')|\ dq'.
$$
\end{theorem}
This result is very important. In fact, any functional order parameter can 
be approximated in the $L^1$ norm through a piecewise constant one. The 
pointwise continuity allows us to deal mostly with piecewise constant order 
parameters.

Now we are ready for the following important definitions.
\begin{definition}
The trial auxiliary function, associated to a given mean field spin glass 
system, as described in Section 2, depending on the functional order 
parameter $x$, is defined as
\begin{equation}\label{trial}
\bar\alpha(\beta,h;x)\equiv \log 2 + 
f(0,h;x,\beta)-\frac{\beta^2}{2}\int_{0}^{1} 
q\, x(q)\,dq.
\end{equation}
\end{definition}
Notice that in this expression the function $f$ appears evaluated at $q=0$, 
and $y=h$, where $h$ is the value of the external magnetic field.

\begin{definition}
The Parisi spontaneously broken replica symmetry solution is defined by
\begin{equation}\label{broken}
\bar\alpha(\beta,h)\equiv \inf_x \bar\alpha(\beta,h;x),
\end{equation}
where the infimum is taken with respect to all functional order parameters 
$x$. 
\end{definition}

Of course, by taking the infimum only with respect to replica symmetric 
order parameters, as in (\ref{replicas}), we would get the replica symmetric 
solution of Sherrington and Kirkpatrick, 
as exploited for example in the sum rules in \cite{Gsum}, 
and \cite{GTquadr}.

The main motivation for the introduction of the quantities given by the 
definitions is the following expected tentative Theorem
\begin{theorem}
\label{dream}
({\bf expected}) In the thermodynamic limit, for the partition function defined in 
(\ref{Z}), we have
$$
\lim_{N\to\infty}N^{-1}E\log Z_{N}(\beta,h,J)=\bar\alpha(\beta,h).
$$
\end{theorem}
Of course, the present technology is  far from being able to give a 
complete rigorous proof. However, in the next Section we will prove that 
$\bar\alpha(\beta,h)$ is at least a rigorous upper bound 
for $N^{-1}E\log Z_{N}(\beta,h,J)$, 
uniformly in $N$.

\section{The main results}

The main results of this paper are summarized in the following
\begin{theorem}
\label{main}
For all values of the inverse temperature $\beta$, and the external 
magnetic field $h$, and for any functional order parameter $x$, the 
following bound holds
$$
N^{-1}E\log Z_{N}(\beta,h,J)\le\bar\alpha(\beta,h;x),
$$
uniformly in $N$, where $\bar\alpha(\beta,h;x)$ is defined in 
(\ref{trial}). Consequently, we have also
$$
N^{-1}E\log Z_{N}(\beta,h,J)\le\bar\alpha(\beta,h),
$$
uniformly in $N$, where $\bar\alpha(\beta,h)$ is defined in 
(\ref{broken}). Moreover, for the thermodynamic limit, we have
$$
\lim_{N\to\infty}N^{-1}E\log Z_{N}(\beta,h,J)\equiv\alpha(\beta,h)\le
\bar\alpha(\beta,h),
$$
and
$$
\lim_{N\to\infty}N^{-1}\log Z_{N}(\beta,h,J)\equiv\alpha(\beta,h)\le
\bar\alpha(\beta,h),
$$
$J$-almost surely.
\end{theorem}

The proof is long, and will be split in a series of intermediate results. 
Consider a generic piecewise constant functional order parameter $x$, as in 
(\ref{xpcws}), and define the auxiliary partition function $\tilde Z$, as 
follows
\begin{eqnarray}
\nonumber
\tilde{Z}_{N}(\beta,h;t;x;J)&\equiv&
\sum_{\sigma_1\dots\sigma_N}\exp\,\left(\beta\sqrt{\frac tN}\sum_{(i,j)}J_{ij}
\sigma_i\sigma_j\right.\\
\label{Ztilde}
&&\left.+\beta h\sum_{i}\sigma_i +\beta\sqrt{1-t}\sum_{a=1}^{K}\sqrt{q_a-q_{a-1}}
\sum_{i}J^{a}_{i}\sigma_i\right).
\end{eqnarray}
Here, we have introduced additional independent centered unit Gaussian 
$J^{a}_{i}$, $a=1,\dots, K$, $i=1,\dots,N$. The interpolating parameter $t$ 
runs in the interval $[0,1]$.

For $a=1,\dots, K$, let us call $E_a$ the average with respect to all 
random variables $J^{a}_{i}$, $i=1,\dots, N$. Analogously, we call $E_0$ 
the average with respect to all $J_{ij}$, and denote by $E$ averages with 
respect to all $J$ random variables.

Now we define recursively the random variables $Z_0,Z_1,\dots,Z_K$
\begin{equation}
\label{Za}
Z_{K}=\tilde{Z}_{N}(\beta,h;t;x;J),\,\,Z_{K-1}^{m_{K}}=E_{K}Z_{K}^{m_{K}}, 
\ldots,\,\,Z_{0}^{m_{1}}=E_{1}Z_{1}^{m_{1}},
\end{equation}   
and the auxiliary function $\tilde\alpha_{N}(t)$
\begin{equation}
\label{atilde}
\tilde\alpha_{N}(t)=\frac{1}{N}E_0 \log Z_{0}.
\end{equation}
Notice that, due to the partial integrations, any $Z_a$ depends only on 
the $J_{ij}$, and on the $J^{b}_{i}$ with $b\le a$, while in $\tilde\alpha$ 
all $J$ noises have been completely averaged out.

The basic motivation for the introduction of $\tilde\alpha$ is given by
\begin{lemma}
\label{inter}
At the extreme values of the interpolating parameter $t$ we have
\begin{eqnarray}\label{t1}
\tilde\alpha_{N}(1)&=&\frac{1}{N}E\log Z_{N}(\beta,h,J),\\
\label{t0}
\tilde\alpha_{N}(0)&=& \log 2 + f(0,h;x,\beta),
\end{eqnarray}
where $f$ is as described in Section 3.
\end{lemma}
The proof is simple. In fact, at $t=1$, the $J^{a}_{i}$ disappear, and 
$\tilde Z$ reduces to $Z$ in (\ref{Z}). On the other hand, at $t=0$, the 
two site couplings $J_{ij}$ disappear, while all effects of the 
$J^{a}_{i}$ factorize with respect to the sites $i$. Therefore, we are 
essentially reduced to a one site problem, and it is immediate to see that 
the averages in (\ref{Za}) reduce to the Gaussian averages necessary for 
the computation of the solution of the antiparabolic problem 
(\ref{antipara}), (\ref{final}), as given  by the repeated 
application of (\ref{solution}), 
with the $f$ function evaluated at $q=0$, 
and $y=h$.

It is clear that now we have to proceed to the calculation of the $t$ 
derivative of $\tilde\alpha_N(t)$. But we need some few additional 
definitions. Introduce the random variables $f_a$, $a=1,\dots,K$,
\begin{equation}
\label{fa}
f_a=\frac{Z_{a}^{m_{a}}}{E_{a}(Z_{a}^{m_{a}})},
\end{equation}
and notice that they depend only on the $J^{b}_{i}$ with $b\le a$, and 
are normalized, $E(f_a)=1$. Moreover, we consider the $t$-dependent state $\omega$ 
associated to the {\it Boltzmannfaktor} in (\ref{Ztilde}), and its 
replicated $\Omega$. A very important role is played by the following 
states $\tilde\omega_a$, and their 
replicated ones $\tilde\Omega_a$, $a=0,\dots,K$, defined as
\begin{equation}\label{omegatildea}
\tilde\omega_K(.)=\omega(.),\,\,\, \tilde\omega_a(.)=E_{a+1}\dots 
E_{K}(f_{a+1}\dots f_{K}\omega(.)).
\end{equation}
Finally, we define the $\med{.}_a$ averages, through a generalization of 
(\ref{medie}),
\begin{equation}\label{mediea}
\med{.}_a=E(f_1 \dots f_a \tilde\Omega_{a}(.)).
\end{equation} 
As it will be clear in the following, the $\med{.}_a$ averages are able, in 
a sense, to concentrate the overlap fluctuations around the value $q_a$.

Now, we have all definitions in order to be able to state the following 
important results.

\begin{theorem}
\label{tdt}
The $t$ derivative of $\tilde\alpha_N(t)$ in (\ref{atilde}) is given by
\begin{eqnarray}
\nonumber
\frac{d}{dt}\tilde\alpha_N(t)&=&-\frac{\beta^2}{4}
\bigl(1-\sum_{a=0}^{K}(m_{a+1}-m_a)q^{2}_{a}\bigr)\\
\label{dt} 
&&-\frac{\beta^2}{4} \sum_{a=0}^{K}(m_{a+1}-m_a)\med{(q_{12}-q_{a})^{2}}_a.
\end{eqnarray}
\end{theorem}

\begin{theorem}
\label{tsum}
For any functional order parameter, of the type given in (\ref{xpcws}), 
the following sum rule holds
\begin{equation}
\label{sum}
\bar\alpha(\beta,h;x)=\frac{1}{N}E\log Z_{N}(\beta,h;J)+
\frac{\beta^2}{4} \sum_{a=0}^{K}(m_{a+1}-m_a)
\int_{0}^{1}\med{(q_{12}-q_{a})^{2}}_{a}(t)\,dt.
\end{equation}
\end{theorem}

Clearly, Theorem \ref{tsum} follows from the previous Theorem \ref{tdt}, by 
integrating with respect to $t$, taking into account the boundary values 
in Lemma 
\ref{inter}, and the definition of $\bar\alpha(\beta,h;x)$ given in Section 3.
Moreover, one should use also  the obvious identity
\begin{equation}
\frac{1}{2} \bigl(1-\sum_{a=0}^{K}(m_{a+1}-m_a)q^{2}_{a}\bigr)=
\int_{0}^{1} q \, x(q) \, dq.
\end{equation}

By taking into account that all terms in the sum rule are nonnegative, 
since $m_{a+1}\ge m_{a}$, we 
can immediately establish the validity of Theorem \ref{main}.

Now we must attack Theorem \ref{tdt}. The proof is straightforward, and 
involves integration by parts with respect to the external noises.
We only sketch the main points. Let us begin with
\begin{lemma}
\label{ldt}
The $t$ derivative of $\tilde\alpha_N(t)$ in (\ref{atilde}) is given by
\begin{equation}
\nonumber
\frac{d}{dt}\tilde\alpha_N(t)=\frac{1}{N}E(f_1 f_2 \dots f_K 
Z_K^{-1}\partial_t Z_K),
\end{equation}
where
\begin{eqnarray}
\nonumber
Z_K^{-1}\partial_t Z_K &=& \tilde Z_N^{-1}\partial_t \tilde Z_N\\
\nonumber
&=&\frac{\beta}{2 \sqrt{t N}}\sum_{(ij)} J_{ij} \omega(\sigma_i \sigma_j) -
\frac{\beta}{2 \sqrt{1-t}}\sum_{a=1}^K \sqrt{q_a - q_{a-1}} \sum_i 
J_i^a\omega(\sigma_i).
\end{eqnarray}
\end{lemma}

The proof is very simple. In fact, from the definitions in (\ref{Za}), we 
have, for $a=0,1,\dots,K-1$,
\begin{equation}
\nonumber
Z_a^{-1}\partial_t Z_a=E_{a+1}(f_{a+1}Z_{a+1}^{-1}\partial_t Z_{a+1}).
\end{equation}
The rest follows from iteration of this formula, and simple calculations.

Clearly, now we have to evaluate
\begin{eqnarray}
\nonumber
E(J_{ij}f_1 f_2 \dots f_K \omega(\sigma_i \sigma_j))&=&\sum_{a=1}^K E(\dots 
\partial_{J_{ij}}f_a\dots\omega(\sigma_i \sigma_j))+
E(f_1 \dots f_K \partial_{J_{ij}}\omega(\sigma_i \sigma_j)),\\
\nonumber
E(J_i^a f_1 f_2 \dots f_K \omega(\sigma_i))&=&\sum_{b=1}^K E(\dots 
\partial_{J_i^a}f_b\dots\omega(\sigma_i ))+
E(f_1 \dots f_K \partial_{J_i^a}\omega(\sigma_i)),
\end{eqnarray}
where we have exploited standard integration by parts on the Gaussian $J$ 
variables.

The following lemma gives all additional information necessary for the 
proof of Theorem~\ref{tdt}.

\begin{lemma}
\label{ldJ}
For the $J$-derivatives we have
\begin{eqnarray}
\label{dijomega}
\partial_{J_{ij}}\omega(\sigma_i \sigma_j)&=&
\beta\sqrt{\frac{t}{N}}(1-\omega^{2}(\sigma_i \sigma_j)),\\
\label{diaomega}
\partial_{J_i^a}\omega(\sigma_i)&=&
\beta\sqrt{1-t}\sqrt{q_a-q_{a-1}}(1-\omega^{2}(\sigma_i)),\\
\label{dijfa}
\partial_{J_{ij}} f_a &=&
\beta\sqrt{\frac{t}{N}} m_a f_a (\tilde\omega_a(\sigma_i \sigma_j)-
\tilde\omega_{a-1}(\sigma_i \sigma_j)),\\
\label{diafb1}
\partial_{J_i^a} f_b &=& 0,\,\,\hbox{if $b<a$},\\
\label{diafa}
\partial_{J_i^a} f_a &=& 
\beta\sqrt{1-t}\sqrt{q_a-q_{a-1}} m_a f_a \tilde\omega_a(\sigma_i),\\   
\label{diafb2}
\partial_{J_i^a} f_b &=&
\beta\sqrt{1-t}\sqrt{q_a-q_{a-1}} m_b f_b (\tilde\omega_b(\sigma_i)-
\tilde\omega_{b-1}(\sigma_i)),\,\,\hbox{if $b>a$}.
\end{eqnarray}
\end{lemma}

The proof of (\ref{dijomega}) and (\ref{diaomega}) is a standard 
calculation. On the other hand, Eq.~(\ref{dijfa}) follows from the definition 
(\ref{fa}) and the easily established
\begin{eqnarray}
\nonumber
\partial_{J_{ij}} Z_{a}^{m_a} &=&
m_a Z_{a}^{m_a} Z_{a}^{-1} \partial_{J_{ij}} Z_a,\\
\nonumber
Z_{a}^{-1} \partial_{J_{ij}} Z_a &=&
E_{a+1}(f_{a+1} Z_{a+1}^{-1} \partial_{J_{ij}} 
Z_{a+1}),\,\,a=1,\dots,K-1,\\
\nonumber
Z_{K}^{-1} \partial_{J_{ij}} Z_K &=&
\tilde Z_{N}^{-1} \partial_{J_{ij}} \tilde Z_N=\beta\sqrt{\frac{t}{N}}
\omega(\sigma_i \sigma_j),\\
\nonumber
Z_{a}^{-1} \partial_{J_{ij}} Z_a &=&
\beta\sqrt{\frac{t}{N}}E_{a+1}(f_{a+1}\dots f_K \omega(\sigma_i \sigma_j))=
\beta\sqrt{\frac{t}{N}}\tilde\omega_a(\sigma_i \sigma_j).
\end{eqnarray}

In the same way, we can establish (\ref{diafb1}), (\ref{diafa}), 
(\ref{diafb2}). But here we have to take into account that $Z_b$ does not 
depend on $J^a_i$ if $b<a$.

A careful combination of all information given by Lemma \ref{ldt} and 
Lemma \ref{ldJ}, finally leads to the proof of  Theorem \ref{tdt}. On the 
other hand, the main Theorem~\ref{main} follows easily from 
Theorem~\ref{tsum}, and the results of \cite{GTthermo}.

\section{Broken replica symmetry bound for the ground state energy}

Let us consider the ground state energy density $-e_N(J,h)$ defined as
\begin{equation}
\label{eN}
-e_N(J,h)\equiv\frac1N\, \inf_{\sigma}H_N(\sigma,h,J)
\nonumber
= -\lim_{\beta \to \infty}\frac{\ln Z_N(\beta,h,J)}{\beta N}.
\end{equation}

By taking the expectation values we also have
\begin{equation}
\label{EeN}
e_N(h)\equiv E(e_N(J,h))=\lim_{\beta\to\infty}\frac{\alpha_N(\beta,h)}{\beta}.
\end{equation}

From the results of the previous Section, we have, for any functional order 
parameter $x$,
\begin{equation}
\label{bound}
\frac{E(\ln Z_N(\beta,h,J))}{\beta N}\le \beta^{-1} \bar\alpha(\beta,h;x),
\end{equation}
uniformly in $N$.

Let us now introduce an arbitrary sequence
\begin{equation}
\label{mbar}
0\le {\bar m}_1 \le {\bar m}_2 \le \dots 
\le {\bar m}_K,
\end{equation}
and the corresponding order parameter $\bar x$, defined as in (\ref{xpcws}), but 
with all $m_a$ replaced by ${\bar m}_a$. Notice that there is no upper 
bound equal to $1$ for 
${\bar m}_K$, and consequently for $\bar x$. However, for sufficiently 
large $\beta$, we definitely have ${\bar m}_K\le \beta$. Therefore, we can 
take in (\ref{bound}) the order parameter $x$ defined by $x(q)={\bar 
x}(q)/\beta$, with $0\le x(q)\le 1$. Then we can easily establish the 
following Lemma.
\begin{lemma}
\label{llim}
In the limit $\beta\to\infty$ we have
\begin{equation}
\lim_{\beta\to\infty} \beta^{-1} \bar\alpha(\beta,h;x)=\tilde\alpha_N(h;{\bar 
x})\equiv {\bar f}(0,h;{\bar x})-\frac12 \int^1_0\,q {\bar x}(q)\, dq,
\end{equation}
\end{lemma}
where the function $\bar f$, with values ${\bar f}(q,y;{\bar x})$ 
satisfies the antiparabolic equation
\begin{equation}
\label{antipara1}
(\partial_q {\bar f})(q,y)+
{1\over2}\bigl({\bar f}^{\prime\prime}(q,y)+
{\bar x}(q){\bar f}^{\prime^2}(q,y)\bigr)=0,
\end{equation}
with final condition
\begin{equation}
\label{final1}
{\bar f}(1,y)=|y|.
\end{equation}
The proof is easy. In fact, the recursive solution for $f$ coming from 
(\ref{solution}), allows to prove immediately
\begin{equation}
\nonumber
\lim_{\beta\to\infty} \beta^{-1} f(q,y;{\bar x}/\beta)={\bar f}(q,y;{\bar 
x}),
\end{equation}
by taking into account the elementary $\lim_{\beta\to\infty} \beta^{-1} 
\log\cosh(\beta y)=|y|$.

Therefore we have established
\begin{theorem}
\label{tbound1}
The following inequalities hold
\begin{eqnarray}
\label{e1}
e_N(h)&\le& \tilde\alpha_N(h;{\bar x}),\\
\label{e2}
e_N(h)&\le& \tilde\alpha_N(h)\equiv \inf_{\bar x}\tilde\alpha_N(h;{\bar x}),\\
\label{e3}
\lim_{N\to\infty}e_N(h)\equiv e_0(h)&\le& \tilde\alpha_N(h;{\bar x}),\\
\label{e4}
e_0(h)&\le& \tilde\alpha_N(h).
\end{eqnarray}
\end{theorem}

A detailed study of the numerical information coming from the variational 
bound of Theorem~\ref{tbound1} will be presented in a forthcoming paper 
\cite{GTe}.  
  
\section{Broken replica symmetry bounds in the p-spin model}

The methods developed in the previous Sections can be easily extended to 
the Derrida p-spin model \cite{D}, \cite{GM}, \cite{Ga}, \cite{Tbook}.  
We give here only a brief sketch. A more 
detailed treatment will be presented elsewhere \cite{GTpspin}.

Now the Hamiltonian contains a term coupling each group made of $p$ spins
\begin{equation}\label{Hp}
H_N(\sigma,h,J)=-\bigl(\frac{p!}{2 N^{p-1}}\bigr)^{\frac12}
\sum_{(i_1,\dots i_p)}J_{i_1\dots i_p}\sigma_{i_1}\dots\sigma_{i_p}
-h\sum_{i}\sigma_i.
\end{equation}

For the sake of simplicity, in the following we consider only the case of 
even $p$. Piecewise constant order parameters are introduced as in 
(\ref{qm}), (\ref{xpcws}), where now we assume $q_K=p/2$. We still introduce 
the function $f$, defined by (\ref{antipara}), with $0\le q\le p/2$, and 
final condition
\begin{equation}
\label{finalp}
f(p/2,y)=\log\cosh(\beta y).
\end{equation}
We also introduce the change of variables $q\to {\bar q}$, 
defined by $2 q = p{\bar q}^{p-1}$, so that, in particular, ${\bar 
q}_K\le1$.  The definitions (\ref{trial}) and
(\ref{broken}) must be modified as follows. 
\begin{definition}
The trial auxiliary function, associated to a given p-spin mean field spin glass 
system, as described before, depending on the functional order 
parameter $x$, is defined as
\begin{equation}\label{trialp}
\bar\alpha(\beta,h;x)\equiv \log 2 + 
f(0,h;x,\beta)-\frac{\beta^2}{2}  \int_{0}^{\frac{p}{2}} 
{\bar q}(q)\, x(q)\,dq.
\end{equation}
\end{definition}

\begin{definition}
The  spontaneously broken replica symmetry solution for the p-spin model 
is defined by
\begin{equation}\label{brokenp}
\bar\alpha(\beta,h)\equiv \inf_x \bar\alpha(\beta,h;x),
\end{equation}
where the infimum is taken with respect to all functional order parameters 
$x$. 
\end{definition}

With the same procedure as described in Section 4, we arrive to the sum 
rule given by
\begin{theorem}
\label{tsump}
In the p-spin model, for any functional order parameter, 
the following sum rule holds
\begin{eqnarray}
\nonumber
\bar\alpha(\beta,h;x)&=&\frac{1}{N}E\log Z_{N}(\beta,h;J)\\
\nonumber
&+&
\frac{\beta^2}{4} \sum_{a=0}^{K}(m_{a+1}-m_a)
\int_{0}^{1}\med{q_{12}^{p}- p q_{12}{\bar q}_{a}^{p-1}+
(p-1) {\bar q}_{a}^{p}}_{a}(t)\,dt\\
\label{sump}
&+&O(\frac1N),
\end{eqnarray}
where $\bar\alpha(\beta,h;x)$ is defined in(\ref{trialp}). 
\end{theorem}

Notice that the terms under the sum are still positive. The $O(\frac1N)$ 
correction is typical of the p-spin models.

From the sum rule we have also
\begin{theorem}
\label{mainp}
In the p-spin model, for any functional order parameter $x$, the 
following bound holds
$$
N^{-1}E\log Z_{N}(\beta,h,J)\le\bar\alpha(\beta,h;x)+O(\frac1N),
$$
where $\bar\alpha(\beta,h;x)$ is defined in 
(\ref{trialp}). Consequently, we have also
$$
N^{-1}E\log Z_{N}(\beta,h,J)\le\bar\alpha(\beta,h)+O(\frac1N),
$$
where $\bar\alpha(\beta,h)$ is defined in 
(\ref{brokenp}). Moreover, for the thermodynamic limit, we have
$$
\lim_{N\to\infty}N^{-1}E\log Z_{N}(\beta,h,J)\equiv\alpha(\beta,h)\le
\bar\alpha(\beta,h),
$$
and
$$
\lim_{N\to\infty}N^{-1}\log Z_{N}(\beta,h,J)\equiv\alpha(\beta,h)\le
\bar\alpha(\beta,h),
$$
$J$-almost surely.
\end{theorem}

We refer to \cite{GTpspin} for a more detailed treatment.

\section{Conclusions and outlook for future developments}

Without exploiting any reference to the zero replica trick, or to the 
cavity method, we have found a way to prove that the true free energy for the 
mean field spin glass model is bounded below by its spontaneously broken symmetry 
expression, given in the frame of the Parisi {\sl Ansatz}. The method 
extends easily to the Derrida p-spin model. The key role is played by the 
auxiliary function $\tilde\alpha_N(t)$, defined in (\ref{atilde}). Our method, 
in its very essence, is a generalization of the mechanical analogy 
introduced in \cite{Gsum}, for the comparison with the replica symmetric 
approximation.

The main open problems are given by the extension of these methods to 
other disordered systems, as for example the mean field neural network models.
Moreover, the sum rules developed here could be taken as the starting 
point to prove that the additional positive terms do really vanish in the 
infinite volume limit. This would prove rigorously the validity of the 
broken replica {\sl Ansatz}.

We plan to report on these problems in future papers.

\vspace{.5cm}
{\bf Acknowledgments}

We gratefully acknowledge useful conversations with
Romeo Brunetti, Enzo Marinari, and 
Giorgio Parisi. The strategy developed in this paper grew out from a 
systematic exploration of interpolation methods, developed in 
collaboration with Fabio Lucio Toninelli.

This work was supported in part by MIUR 
(Italian Minister of Instruction, University and Research), 
and by INFN (Italian National Institute for Nuclear Physics).

\end{document}